%% file: _main.tex
\input{_constants}
\arxiv  

\pdfoutput=1
\documentclass[10pt,twocolumn,letterpaper]{article}
\input{cvpr_header}
\begin{document}
\title{Spatiotemporal Modeling Encounters 3D Medical Image Analysis: \\Slice-Shift UNet with Multi-View Fusion}
\author{
  \begin{tabular}[t]{c}
    Cynthia I. Ugwu \and Sofia Casarin \and Oswald Lanz \\
  \end{tabular} \\
  \begin{tabular}[t]{c}
    Free University of Bozen-Bolzano \\
    {\tt\small \{cugwu, scasarin, oswald.lanz\}@unibz.it}
  \end{tabular}
}

\maketitle

\input{00_abstract}
\input{01_intro}
\input{02_related}

\input{03_method}

\input{10_conclusion}

\subsection*{Acknowledgements} The research presented in this paper was partially funded by \textit{Covision Lab} and the Italian \textit{National Operative Program}, budget for ”Research and Innovation” 2014-2020 under Action IV.5 ”Doctorates on green topics”.

{\small
\bibliographystyle{ieee_fullname}
\bibliography{11_references}
}

\ifarxiv \clearpage \input{12_appendix} \fi

\end{document}

%% file: _constants.tex
\def\cvprPaperID{0000}
\def\confName{CVPR}
\def\confYear{2023}

\newif\ifreview 
\newif\ifarxiv \newcommand{\arxiv}{\arxivtrue}
\newif\ifcamera 
\newif\ifrebuttal 

%% file: cvpr_header.tex
\ifreview \usepackage[review]{cvpr} \fi
\ifarxiv \usepackage[pagenumbers]{cvpr} \fi
\ifrebuttal \usepackage[rebuttal]{cvpr} \fi
\ifcamera \usepackage{cvpr} \fi

\usepackage{graphicx}
\usepackage{amsmath}
\usepackage{amssymb}
\usepackage{booktabs}

\input{_macros}  

\usepackage{xr-hyper}

\makeatletter
\newcommand*{\addFileDependency}[1]{
  \typeout{(#1)}
  \@addtofilelist{#1}
  \IfFileExists{#1}{}{\typeout{No file #1.}}
}

\makeatother

\usepackage[pagebackref,breaklinks,colorlinks]{hyperref}
\usepackage[capitalize]{cleveref}
\crefname{section}{Sec.}{Secs.}
\crefname{table}{Table}{Tables}
\crefname{figure}{Fig.}{Figs.}

%% file: _macros.tex
\usepackage{times}
\usepackage{microtype}
\usepackage{epsfig}
\usepackage[table,xcdraw]{xcolor}
\usepackage{caption}
\usepackage{float}
\usepackage{placeins}
\usepackage{color, colortbl}
\usepackage{stfloats}
\usepackage{enumitem}
\usepackage{tabularx}
\usepackage{xstring}
\usepackage{multirow}
\usepackage{xspace}
\usepackage{url}
\usepackage{subcaption}
\usepackage{xcolor}
\usepackage[hang,flushmargin]{footmisc}

\usepackage{multirow}
\usepackage{adjustbox}

\ifcamera \usepackage[accsupp]{axessibility} \fi

% ...

\ifarxiv  \fi

\newcommand{\R}[1]{{%
    \textbf{%
        \ifstrequal{#1}{1}{\textcolor{red}{R#1}}{%
        \ifstrequal{#1}{2}{\textcolor{blue}{R#1}}{%
        \ifstrequal{#1}{3}{\textcolor{magenta}{R#1}}{%
        \ifstrequal{#1}{4}{\textcolor{teal}{R#1}}{%
                           \textcolor{cyan}{R#1}%
        }}}}%
    }%
}}

%% file: 00_abstract.tex
\begin{abstract}
    As a fundamental part of computational healthcare, Computer Tomography (CT) and Magnetic Resonance Imaging (MRI) provide volumetric data, making the development of algorithms for 3D image analysis a necessity. Despite being computationally cheap, 2D Convolutional Neural Networks can only extract spatial information. In contrast, 3D CNNs can extract three-dimensional features, but they have higher computational costs and latency, which is a limitation for clinical practice that requires fast and efficient models. 
    Inspired by the field of video action recognition we propose a new 2D-based model dubbed Slice SHift UNet (SSH-UNet) which encodes three-dimensional features at 2D CNN's complexity. More precisely multi-view features are collaboratively learned by performing 2D convolutions along the three orthogonal planes of a volume and imposing a weights-sharing mechanism. The third dimension, which is neglected by the 2D convolution, is reincorporated by shifting a portion of the feature maps along the slices' axis. The effectiveness of our approach is validated in Multi-Modality Abdominal Multi-Organ Segmentation (AMOS) and Multi-Atlas Labeling Beyond the Cranial Vault (BTCV) datasets, showing that SSH-UNet is more efficient while on par in performance with state-of-the-art architectures.
\end{abstract}

%% file: 01_intro.tex
\section{Introduction}
\label{sec:intro}
Identifying organs through semantic segmentation is a crucial step in several clinical workflows, including diagnosis, intervention, therapy planning, treatment delivery, and tumour growth monitoring. However, the volumetric data generated by medical acquisition systems, such as Computer Tomography (CT), Magnetic Resonance Imaging (MRI), or Ultrasound, can make the segmentation task labour-intensive and time-consuming. For instance, a single 3D CT scan can contain hundreds of 2D slices (images). Therefore, developing robust and accurate automatic segmentation tools is a fundamental necessity in medical image analysis~\cite{Sun_2020, Tang_2019}.
\begin{figure}[t]
\begin{center}
   \includegraphics[width=1.0\linewidth]{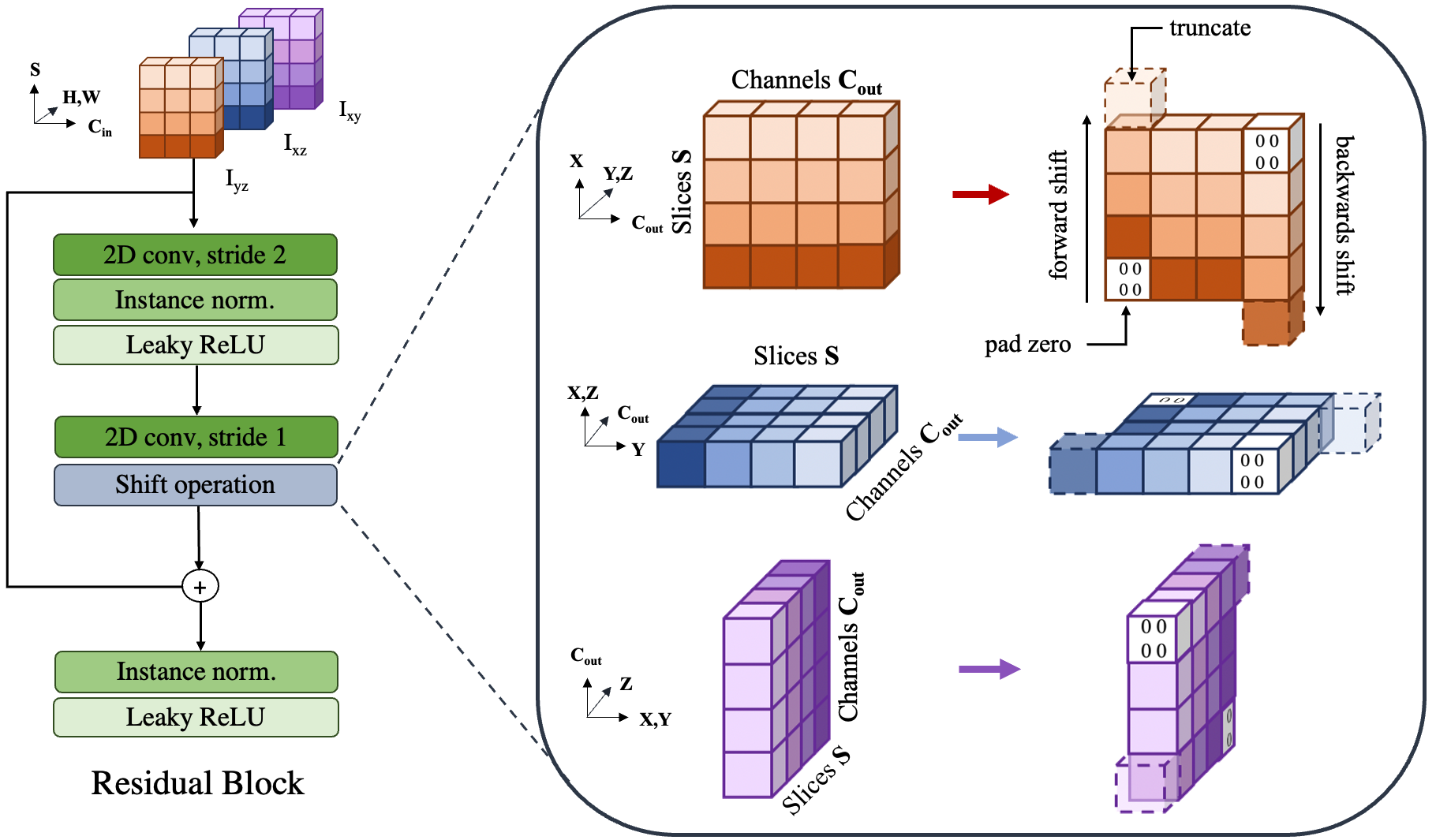}
\end{center}
   \caption{Overview of the proposed framework. An SSH-UNet's layer is a Residual Block receiving an input tensor with dimension $(3B, C_{in}, S, H, W)$, where $3B$ is the concatenation on the batch $B$ of the features from the three orthogonal planes of the CT volume ($I_{xy}, I_{yz}, I_{xz}$). Spatial features are extracted by a 2D convolution from $(H,W)$, and are then shifted forward and backwards along slices' axis $S$. The operation is performed for each tensor $I_{xy}, I_{yz}, I_{xz}$ in the batch. Since we are interested in how features are mixed, we represented the shift with the three axes explicitly depicting the slices and channels dimensions while the spatial dimensions $H$ and $W$ are condensed on a single axis.
} 
\label{fig:pipeline}
\end{figure}
With the advent of deep learning, Convolutional Neural Networks (CNNs) have proved to be extremely effective at solving vision tasks due to their powerful representation learning capabilities. In particular, "U-shaped" encoder-decoder architectures have achieved state-of-the-art results in various medical semantic segmentation tasks~\cite{chen2019s3d,futrega2022optimized,Isensee_2021}. More recently, Vision Transformers (ViT)~\cite{dosovitskiy2021an} have achieved comparable results to CNN-based methods, and as a result, many transformer-based models have been proposed for both 2D and 3D medical image segmentation~\cite{Xie_2021,cao2021swin,chen2021transunet}.
Although 3D CNNs are designed to learn three-dimensional features, they require higher computation costs, resulting in higher inference latency compared to 2D CNNs. Besides, the large number of parameters may result in a higher risk of overfitting, especially when encountering small datasets~\cite{zhang2022bridging}. This is very common in the medical field as it is challenging to collect 3D medical datasets due to accessibility issues for ethical reasons, and limited time and budget for annotations.
To process volumetric data more efficiently, two main strategies can be used. The first one is cutting the volume into slices and training 2D CNNs to segment each slice separately ~\cite{cao2021swin,chen2021transunet}. Despite the computational efficiency, as the information between adjacent slices is neglected, it leads to segmentation results that are prone to discontinuity in 3D space~\cite{zhang2022bridging}. The second is using 2.5D segmentation methods (or pseudo-3D methods). A very common 2.5D strategy is "multi-view fusion" where three 2D CNNs are trained on the sagittal, coronal, and axial planes separately~\cite{yun2019improvement}, after that, the segmentation results from each plane are fused to get the final result. 

In this work, we propose a bi-dimensional UNet, for segmentation on volumetric medical data that extracts multi-view and multi-slice information thanks to a Slice SHift mechanism (SSH-UNet). To extract multi-view features as in~\cite{li2019collaborative} we impose weight sharing between the 2D convolution that processes the slices from the three orthogonal planes. While shifting is a well-established technique in video processing, we wondered if it could also be transferred to volumetric data since there is no inherent preferential direction like time in videos. As shown in Figure~\ref{fig:pipeline} intra-slice features are extracted by shifting a portion of the feature maps along the slices' axis following the work in~\cite{Lin_2019}. SSH-UNet is evaluated on two publicly available benchmark datasets the Multi-Modality Abdominal Multi-Organ Segmentation (AMOS)~\cite{ji2022amos} and Multi-Atlas Labeling Beyond the Cranial Vault (BTCV)~\cite{BTCV_orig}. To the extent of our knowledge, no previous work in the medical field has explored the combination of shifting and shared weights across multiple views within a single model. 

To be more specific, the contributions of our work are as follows:
\begin{itemize}
    \item We propose the first network that repurposes the spatiotemporal modelling in video tasks to segment medical data. By interpreting the slices' axis as the time, we solve the problem of 2D CNNs that neglect information between adjacent slices by shifting a portion of the feature maps along the slices' axis.
    \item We revisit and extend the 2.5D multi-view fusion method by processing slices from the three orthogonal planes of a volume using a 2D UNet with shared weights rather than three separate networks, allowing multi-view features to be learned collaboratively while maintaining a light computational cost.
    \item We instantiate these ideas into the Slice-Shift (SSH) layer, a 2D convolution layer operating on 3D tensors. We validate the effectiveness of the proposed framework by training a UNet built of SSH layers on two publicly available benchmark datasets, AMOS and BTCV, showing that our approach with the same model complexity as 2D CNNs achieves the same performance as a fully 3D network with a similar architecture and can achieve comparable results with other popular state-of-the-art approaches with less than $1/5$ of parameters.
\end{itemize}
Our code will be released to facilitate follow-up research.

%% file: 02_related.tex
\section{Related Work}
\label{sec:related}

\subsection{Segmentation on medical data with U-Net}
\label{subsec:unet-like}
U-Net~\cite{Ronneberger_2015} was proposed for biomedical image segmentation back in 2015. Afterwards, a new class of models was developed based on U-Net-like architectures which established the state-of-the-art in segmentation. One promising approach was proposed by Isensee \etal in~\cite{Isensee_2021}, where nnU-Net was introduced. nnU-Net is a deep learning-based segmentation method that automatically configures itself for any new task. 
Its performance is not attained through a new architecture (thus the name nnU-Net, 'no new net'), as it only comprises minor modifications to the original U-Net. Rather, it automates the complicated process of manually configuring the method. 
Hatamizadeh \etal reformulate in~\cite{Hatamizadeh_2022} the task of volumetric medical image segmentation as a sequence-to-sequence prediction problem by leveraging the power of self-attention and transformers architectures. They introduce a novel architecture, dubbed as UNEt TRansformers (UNETR), that utilizes a transformer as the encoder. The extracted representations are merged with a CNN-based decoder via skip connections at multiple resolutions. The ensemble of UNETR models has shown promising results on the BTCV dataset. Tang \etal introduced in~\cite{Tang_2022} a novel 3D transformer-based model dubbed Swin UNEt TRansformers (Swin UNETR). Swin UNETR comprises a Swin Transformer~\cite{Liu_2021} encoder and a CNN-based decoder. The transformer encoder is pre-trained with tailored, self-supervised tasks over 5,050 images. Overall, the ensemble of 20 Swin UNETR models achieved at the time of publication the top-ranking performance on the BTCV challenge, showing distinct improvements for the segmentation of organs that are smaller in size. 

\subsection{Video action recognition}
\label{subsec:video}
Spatio-temporal representation learning refers to the process of learning meaningful representations of both spatial and temporal information in a given dataset. In computer vision, this is particularly important for tasks such as video analysis and action recognition, where the goal is to accurately model the spatial and temporal evolution of objects and subjects over time. In particular, video action recognition has received increasing attention due to its potential applications such as video surveillance, human-computer interaction, and social video recommendation. This field presents however a fundamental challenge due to the space-time nature of the data. For years many efforts were made to trade off between temporal modelling and computation (\cite{Wang_2016}, \cite{Luo_2019}, \cite{Lin_2019}, \cite{Sudhakaran_2020}). Conventional 2D CNNs are computationally cheap but cannot capture temporal relationships. Since a video can be seen as a temporally dense sampled sequence of images, expanding the 2D convolution operation to 3D convolution is an intuitive approach to spatiotemporal feature learning. While 3D CNN-based methods can achieve strong results, they require significant computational resources. Lin \etal proposed in their work~\cite{Lin_2019} a Temporal Shift Module (TSM) that can achieve the performance of 3D CNN but maintain 2D CNN's complexity. TSM shifts a fixed amount of the channels along the temporal dimension, facilitating information exchange among neighbouring frames yielding a 2D CNN that can learn spatiotemporal features. Li \etal \cite{li2019collaborative} propose an operation that encodes spatiotemporal features by imposing a weight-sharing constraint. In particular, they perform 2D convolution by sharing the convolution kernels of three orthogonal views of a video, allowing multi-view features to be learned collaboratively.

%% file: 03_method.tex
\section{Slice-Shift UNet}
\label{sec:method}
\begin{figure*}
\begin{center}
\includegraphics[width=1.0\linewidth]{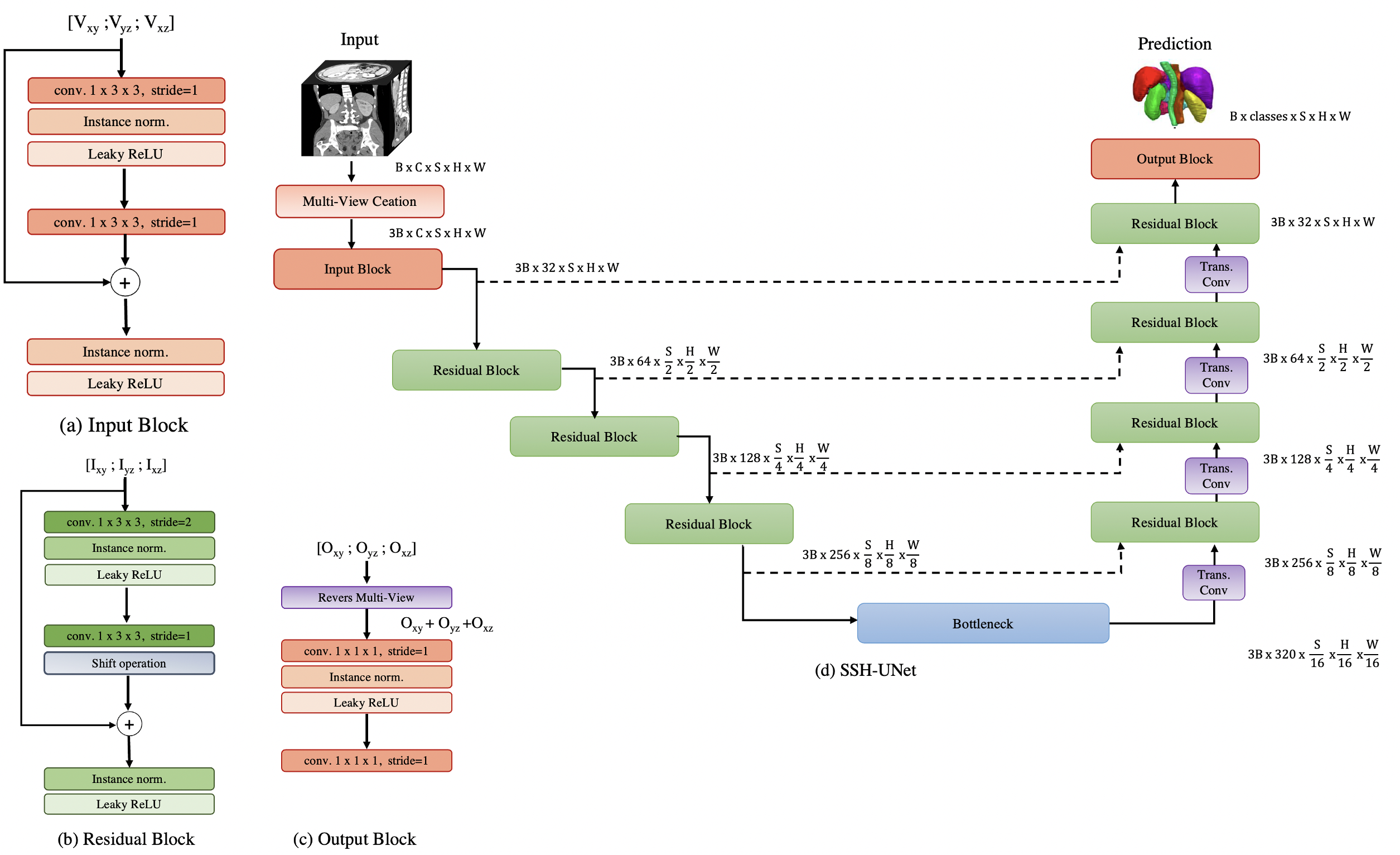}
\end{center}
   \caption{Detailed architecture and components of our proposed SSH-UNet. (a) Input block of our network that processes the concatenation in the batch of the three created views, (b) Residual block that can extract multi-view and multi-slice features thanks to 2D convolutions with shared weights and slice shift mechanism. (c) Output block where multi-view creation is reversed and multi-view fusion is performed to obtain the final segmentation mask. d) Overview of SSH-UNet architecture.}
\label{fig:sshunet}
\end{figure*}

We based our model design on the UNet architecture proposed by Isensee \etal in~\cite{Isensee_2021} and optimized by Futrega \etal in~\cite{futrega2022optimized}. SSH-UNet, whose detailed illustration is found in Figure~\ref{fig:sshunet}, is a CNN-based architecture designed to capture the global connections between multi-plane (axial, coronal, and sagittal) and multi-slice images. This is obtained through weight sharing and by shifting the feature maps along the slices' axis. The overall framework is characterized by: 1) 2D residual blocks used to extract spatial features from the slices of the input volume, 2) slice shifting to incorporate information between adjacent slices neglected by 2D convolutions, and 3) a multi-view fusion block to obtain the final segmentation predictions from the three orthogonal planes.
\subsection{2D residual block}
\label{subsec:2dconv}
Let us assume that the input to the encoder is a sub-volume $\textbf{V}\in \mathbb{R}^{C_{in}\times S\times H\times W}$, with $C_{in}$ channels and patch resolution of $(S, H, W)$.
$\textbf{V}$ lies in the Euclidean space, thus it has three mutually perpendicular coordinate axes $\textit{x}, \textit{y}$ and $\textit{z}$ and three mutually perpendicular coordinate planes: $\textit{xy-plane}, \textit{yz-plane}$ and $\textit{xz-plane}$. For clarity we use the following notation $\textbf{V}\in \mathbb{R}^{C_{in}\times X\times Y\times Z}$. 
We modify the input tensor by placing the plane of interest on the last two dimensions. More precisely from $\textbf{V}$ we generate three volumes $\textbf{V}_{xy}$, $\textbf{V}_{yz}$ and $\textbf{V}_{xz}$, as below:
\begin{align}
     \textbf{V}_{xy}: \mathbb{R}^{C_{in}\times X\times Y\times Z} \rightarrow \mathbb{R}^{C_{in}\times Z\times X\times Y} \nonumber \label{eq:reshape} \\
     \textbf{V}_{yz}: \mathbb{R}^{C_{in}\times X\times Y\times Z} \rightarrow \mathbb{R}^{C_{in}\times X\times Y\times Z}\\ \nonumber
     \textbf{V}_{xz}: \mathbb{R}^{C_{in}\times X\times Y\times Z} \rightarrow \mathbb{R}^{C_{in}\times Y\times X\times Z} \nonumber 
\end{align}
The three volumes are stored in the batch dimension obtaining the final input $\textbf{I}$:
\begin{equation}
    \textbf{I}=[\textbf{V}_{xy}; \textbf{V}_{yz}; \textbf{V}_{xz}].
\label{eq:input}
\end{equation}
We apply 2D convolution with a kernel size of $1\times k\times k$ extracting spatial features from the three orthogonal planes stored in $\textbf{I}$. Methods like~\cite{prasoon2013deep,roth2014new} treat images from $xy$, $yz$, and $xz$ planes as three channels of 2D images. This is empirically effective and memory efficient, but the weakness of the approach is that the three channels are not spatially aligned~\cite{jang2022m3t}, which is why we chose to concatenate the three views in the batch leaving the network to learn multi-view features through weights shearing.

Overall, our residual block is composed of two convolutional layers with kernel size $1\times 3\times 3$ followed by instance normalization and LeakyReLu activation. A residual skip connects the input of the block with the output of the second convolution.

\subsection{Slice shifting} 
Given a volume $\textbf{V}\in \mathbb{R}^{C\times S\times H\times W}$ perceived as a sequence of $S$ images (or slices) with resolution $(H, W)$,  when applying 2D convolution, we do not extract features between adjacent slices. In SSH-UNet we apply a shift operation to re-integrate the third dimension and mingle the information in neighbouring slices.
The intuition behind the shift operation adapted from \cite{Lin_2019} is the following: if we consider a 1-D convolution with kernel size 3 and weights $W=(w_1, w_2, w_3)$, and a 1D input tensor $X$, then the convolution operation can be written as $Y_i=w_1X_{i-1} + w_2X_i + w_3X_{i+1}$. The operation can be decoupled as a \textit{shift} and \textit{multiply-accumulate}, where $X$ is shifted by -1, 0, +1 and multiplied by $(w_1, w_2, w_3)$ respectively. The \textit{shift} operation is:
\begin{equation}
    X^{-1}_i=X_{i-1} ,\quad X^0_i=X_i, \quad X^{+1}_i=X_{i+1} 
\end{equation}
which can be conducted separately from multiplication.
The \textit{multiply-accumulate} operation is:
\begin{equation}
    Y=w_1X^{-1} +  w_2X^{0} + w_3X^{+1}
\end{equation}
that in our case is computed by the previously mentioned 2D convolution. The shift operation does not introduce any extra computational cost to the 2D CNN model. The overall framework is described in Figure \ref{fig:pipeline} where an intermediate residual layer of SSH-Unet with $C_{in}$ input channels and $C_{out}$ output channels is depicted. The slices' axis $S$ change based on the plane we are considering: axis $X$, $Y$, $Z$ for the axial, sagittal, and coronal planes respectively. The feature maps of the different slices are denoted with different shades of colours in each row. Along the slices' axis, we shift part of the channels forward and backwards by +1 and -1 leaving the rest un-shifted. We shift a proportion of $1/4$ of the channels forward and $1/4$ backwards.

\subsection{Multi-view fusion}
The last Residual block of the decoder gives as output the tensor
$\textbf{O}=[\textbf{O}_{xy}; \textbf{O}_{yz};\textbf{O}_{xz}]$. The final segmentation mask is obtained by fusing the three output tensors stored in the batch of $\textbf{O}$. In the first place, the operation computed in Eq. \ref{eq:reshape} is reversed
\begin{align}
    \textbf{O}_{xy}: \mathbb{R}^{C\times Z\times X\times Y} \rightarrow \mathbb{R}^{C\times X\times Y\times Z} \nonumber \\
    \textbf{O}_{yz}: \mathbb{R}^{C\times X\times Y\times Z} \rightarrow \mathbb{R}^{C\times X\times Y\times Z} \nonumber \\
    \textbf{O}_{xz}: \mathbb{R}^{C\times Y\times X\times Z} \rightarrow \mathbb{R}^{C\times X\times Y\times Z} \nonumber 
\label{eq:reverse}
\end{align}
ensuring that information isn't wrongly mixed when tensors are fused. After, $\textbf{O}_{xy}, \textbf{O}_{xz},$ and $\textbf{O}_{yz}$ are summed followed by two convolutions with kernel size $1\times 1\times 1$ generating the final segmentation mask.

\section{Experiments}
\label{sec:experiments}
\subsection{Datasets} 
\textbf{AMOS:} the Multi-Modal Abdominal Multi-Organ Segmentation dataset~\cite{ji2022amos} was introduced as part of the MICCAI 2022 challenge. AMOS is a large-scale, diverse, clinical dataset for abdominal organ segmentation that provides 500 CT and 100 MRI scans accompanied by voxel-level annotations for 15 organs. The data were collected from Longgang District Central Hospital (SZ, China). With over 74k annotated slices AMOS is $\times20$ larger than BTCV~\cite{BTCV_orig} dataset (3.6K annotated slices). For our experiment, we use the AMOS-CT subset where all the 500 CT scans are interpolated into the isotropic voxel spacing of $1.0\times1.0\times1.0$ $mm^3$. Following~\cite{ji2022amos} we first truncate the HU values between $[-991, 362]$ and normalize to $[0,1]$. Data augmentation of random flip, rotation, intensities scaling, and shifting are used with probabilities set to 0.2, 0.2, 0.5, and 0.5 respectively. The multi-organ segmentation problem is formulated as a 16-class segmentation task with 1-channel input.

\textbf{BTCV:} For the ablation analysis (Section \ref{sec:ablation}), we utilize the popular Multi-Atlas Labeling  Beyond the Cranial Vault dataset~\cite{BTCV_orig}. BTCV contains 30 subjects with abdominal CT scans where 13 organs are annotated by interpreters under the supervision of radiologists at Vanderbilt University Medical Center. All CT scans were interpolated into the isotropic voxel spacing of $1.0\times 1.0\times 1.0$ $mm^3$ as a pre-processing step. The intensity was truncated between $[-175, 250]$ and normalized to $[0,1]$. We used the same data augmentation implemented in AMOS.

\subsection{Implementation details} 
The network architecture was created using as baseline DynUNet class from MONAI\footnote {\url{https://monai.io/}}. We extended the original class by inserting the slice shifting in its building blocks and by adding our Multi-View Creation step and Multi-View Fusion Block. For a fair comparison the results in Table \ref{tab:amos} are obtained by training for 1000 epochs using SGD optimizer with a momentum of 0.99, warm-up cosine scheduler for 50 iterations, an initial learning rate of 0.01,  and a batch size of 2, recreating the same training condition of the benchmark created in~\cite{ji2022amos}. Following the official AMOC-CT challenge data split we used 200 CT scans for training and 100 CT scans for the validation set. With the BTCV dataset, we trained for 5000 epochs and stopped the training after 1000 epochs if the validation accuracy did not improve. An AdamW optimizer with a warm-up cosine scheduler was used for 50 iterations, batch size 2, an initial learning rate of 4$e$-4, momentum of 0.9, and decay rate of 1$e$-5. We used 24 CT for training and 6 CT for testing. 

Each training was conducted with a patch resolution of $96\times 96\times 96$ on an NVIDIA A100.

\subsection{Evaluation metric} 
We used the Dice Similarity Coefficient (DSC) and the Normalized Surface Dice (NSD)~\cite{nikolov2018deep} metric to evaluate the segmentation accuracy in our experiments. While DSC measures the overlap between two volumes, the NSD score provides information on the segmentation quality for the boundaries. Given the ground truth $Y$ and the prediction $\hat{Y}$ for each voxel $i$ the Dice score is defined as:
\begin{equation}
    Dice=2\cdot \frac{\sum_{i=1}^IY_i\hat{Y_i}}{\sum_{i=1}^IY_i + \sum_{i=1}^I\hat{Y_i}}.
\end{equation}
Using the above two metrics, we calculate category-wise performance. The DSC used to gauge model performance, ranges from 0 to 1, where 1 corresponds to a pixel-perfect match between the deep learning model output $\hat{Y}$ and ground truth annotation $Y$. The NSD is used to determine which fraction of a segmentation boundary is correctly predicted with values ranging between 0 and 1.
\section{Results}
\label{sec:results}
We compare our model with six state-of-the-art medical segmentation methods present in the benchmark in~\cite{ji2022amos} where Yuanfeng and his colleagues, for the training stage, randomly cropped sub-volumes of size $64\times 160\times 160$; we rather cropped sub-volumes of size $96\times 96\times 96$ as input for our network, due to the multi-view creation, described in Section \ref{subsec:2dconv}, that requires an isotropic volume size. The implementation of the state-of-the-art methods can be found in: UNet\footnote{\url{https://github.com/MIC-DKFZ/nnUNet/tree/master}},  VNet\footnote{\label{monai_nets}\url{https://github.com/Project-MONAI/MONAI/tree/dev/monai/networks/nets}}, CoTr\footnote{\url{https://github.com/YtongXie/CoTr/tree/main/CoTr_package/CoTr}}, nnFormer\footnote{\url{https://github.com/282857341/nnFormer/tree/main/nnformer}}, UNetr\footref{monai_nets}, Swin-UNetr\footref{monai_nets}. 

\begin{figure}[t]
\begin{center}
\includegraphics[width=1.0\linewidth]{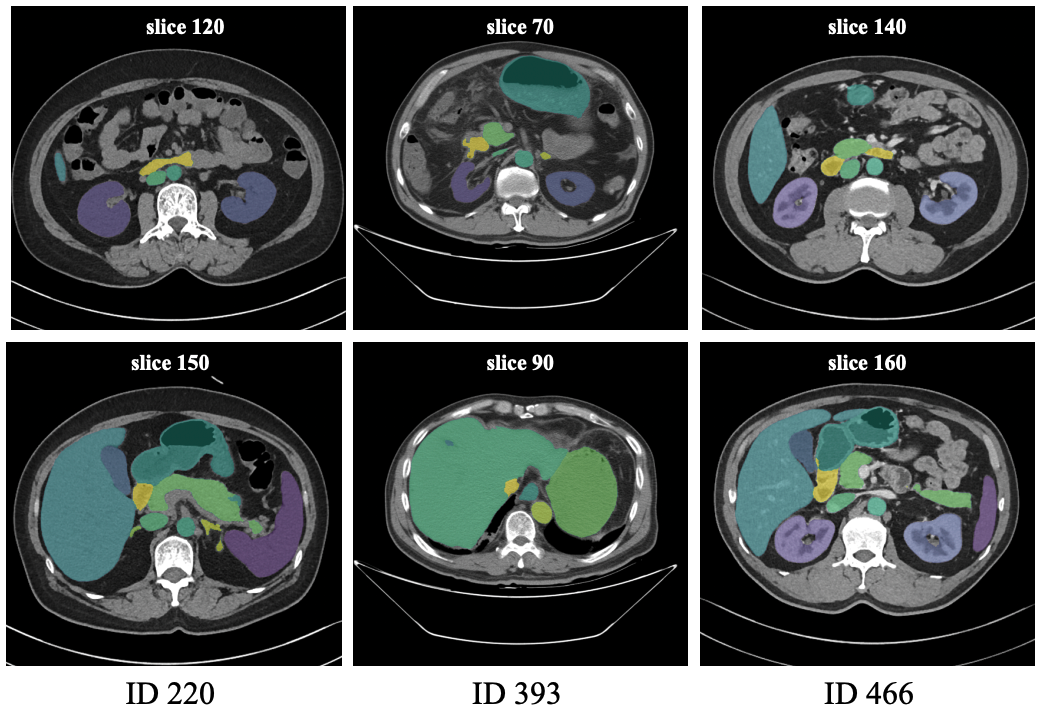}
\end{center}
   \caption{In this qualitative visualization we can see the prediction of SSH-UNet for three samples, identified by their ID number, from the AMOS-CT test set.}
\label{fig:amos_results}
\end{figure}

The class-wise Dice scores on the AMOS-CT validation set are shown in Table \ref{tab:amos}. By training with $96\times 96\times 96$ patches, we achieve an overall accuracy of \textbf{87.28\%} gaining the second position in the benchmark right after UNet~\cite{Isensee_2021}, trained with $64\times 160\times 160$, that indeed outperforms SSH-UNet with +1.6\% gain in accuracy. However, our model has almost -80\% of parameters. Comparing SSH-UNet with Swin UNETR~\cite{Tang_2022} (previously ranked first on MSD~\cite{msd} and BTCV leaderboards) our model offers a substantial improvement in segmenting: right kidney +2.2\%, gallbladder +5.8\%, liver +1.7\%, stomach +3.4\%, and prostate/uterus +4.2\%. In Table \ref{tab:amos_test} the overall results from the AMOS-CT test benchmark are shown. SSH-UNet also confirmed its second position in the test set with an average DSC of \textbf{87.75\%} and NSD of \textbf{77.16\%}. The class-wise DSC and NSD can be found in Table \ref{tab:class_test_amos}, while Figure \ref{fig:amos_results} shows some representative samples of our predictions.

\begin{table*}[t]
  \begin{center}
  \begin{adjustbox}{max width=\textwidth}   
\begin{tabular}{l|rrrrrrrrrrrrrrr|r}
\hline
\multirow{2}{*}{Models} &
      \multicolumn{16}{c}{Categorical DSC(\%) $\uparrow$} \\
 & SPL& RKI &LKI &GBL &ESO &LIV &STO &AOR& IVC &PAN &RAG &LAG &DUO& BLA &PRO/UTE & \textbf{Avg.} \\
\hline
UNet~\cite{Isensee_2021}& \textbf{96.31}& 95.29 & \textbf{96.28}& 81.53& \textbf{85.72} & 97.05 & 90.77 & \textbf{95.37} & \textbf{91.53} & \textbf{87.39} & \textbf{79.83} & \textbf{81.12} & \textbf{82.56} & \textbf{88.42} & \textbf{83.81} & \textbf{88.87} \\
 VNet~\cite{milletari2016v}&  94.21 &91.86 & 92.65 &70.25 & 79.04 & 94.65 & 84.79 & 92.96 & 87.4 & 80.5 & 72.62 &73.19 &71.69 & 77.02 & 66.62& 81.96\\
 CoTr~\cite{Xie_2021}&  91.09 & 87.18& 86.36& 60.47&80.9 &91.61 & 80.09& 93.66& 87.72 & 76.32 & 73.68& 71.74&67.98& 67.38& 40.84 & 77.13\\
nnFormer~\cite{zhou2021nnformer}& 95.91& 93.51& 94.8& 78.47& 81.09& 95.89& 89.4& 94.16& 88.25& 85.0& 75.04 &75.92& 78.45& 83.91& 74.58& 85.63\\
UNETR~\cite{Hatamizadeh_2022}&  92.68& 88.46& 90.57& 66.5& 73.31& 94.11& 78.73& 91.37& 83.99& 74.49& 68.15& 65.28& 62.35& 77.44 &67.52& 78.33\\
Swin-UNETR~\cite{Tang_2022}& 95.49 &93.82 &94.47 &77.34 &83.05 &95.95 &88.94 &94.66 &89.58 &84.91 &77.2 &78.35 &78.59 &85.79 &77.39& 86.37\\
\hline
SSH-UNet & 95.77 & \textbf{96.01} & 94.29 & \textbf{83.12} & 81.81 & \textbf{97.60} & \textbf{92.32} & 94.34& 88.42& 85.36& 76.43& 76.36& 77.79& 87.99& 81.54 & 87.28 \\
\hline
\end{tabular}
\end{adjustbox}
\end{center}
\caption{The class-wise Dice score on the validation set of AMOS-CT. We compare SSH-UNet with the official benchmark on~\cite{ji2022amos}. Note: spleen (SPL), right kidney (RKI), left kidney (LKI), gallbladder (gbl), esophagus (ESO), liver (LIV), stomach (STO), aorta (AOR), inferior vena cava (IVC), pancreas (PAN), right adrenal gland (RAG), left adrenal gland (LAG), duodenum (DUO), bladder (BLA), prostate/uterus (PRO/UTE). The best results are highlighted in bold. SSH-UNet outperforms UNet on four organs and has an average segmentation accuracy 1.6\% inferior with respect to 3D UNet, but has 20\% of UNet's parameters, as shown in Figure~\ref{fig:complexity}.}
\label{tab:amos}
\end{table*}

\begin{table*}[t]
  \begin{center}
  \begin{adjustbox}{max width=\textwidth}   
\begin{tabular}{l|rrrrrrrrrrrrrrr|r}
    \hline
\multirow{2}{*}{SSH-UNet} &
      \multicolumn{15}{c}{CT-Test} \\
    & SPL& RKI &LKI &GBL &ESO &LIV &STO &AOR& IVC &PAN &RAG &LAG &DUO& BLA &PRO/UTE & \textbf{Avg.} \\
    \hline
    DSC & 95.41 & 96.17 & 94.63 & 82.65 & 83.09 & 97.80 & 92.45 & 94.26 & 90.12 & 85.29 & 77.35 & 79.40 & 78.60 & 89.12 & 79.98 & 87.75 \\
    NSD & 88.97 & 89.06 & 86.54 & 72.55 & 73.39 & 85.01 & 76.86 & 87.77 & 75.70 & 68.30 & 81.01 & 80.63 & 62.24 & 75.15 & 54.29 & 77.16\\ 
    \hline
\end{tabular}
\end{adjustbox}
\end{center}
\caption{The class-wise Dice score (DSC) and the Normalized Surface Distance (NSD) of SSH-UNet on the AMOS-CT test.}
\label{tab:class_test_amos}
\end{table*}

\begin{table}[t]
  \begin{center}
    {\small{
\begin{tabular}{l|rr}
\hline
\multirow{2}{*}{Models} &
      \multicolumn{2}{c}{CT-Test} \\
       &mDSC(\%)& mNSD(\%) \\
\hline
UNet~\cite{Isensee_2021}& \textbf{89.04} & \textbf{78.32} \\
 VNet~\cite{milletari2016v}& 82.92 & 67.56  \\
 CoTr~\cite{Xie_2021}&  80.86 & 66.31\\
nnFormer~\cite{zhou2021nnformer}& 85.61 & 72.48\\
UNETR~\cite{Hatamizadeh_2022}& 79.43 & 60.84 \\
Swin-UNETR~\cite{Tang_2022}& 86.32 & 73.83 \\
\hline
SSH-UNet & 87.75 & 77.16 \\
\hline
\end{tabular}
}}
\end{center}
\caption{Overall results of six state-of-the-art methods taken from the official AMOS-CT test benchmark in~\cite{ji2022amos} and SSH-UNet.}
\label{tab:amos_test}
\end{table}

In Table \ref{tab:per_organ_kfold} we can see the results of 5-fold cross-validation on the BTCV dataset. On average our model is able to reach \textbf{84.35\%} of accuracy without the help of any ensemble. From the table, we can observe that the fourth-fold segmentation of the spleen shows a significant drop in performance. The gallbladder and adrenal glands are segmented poorly by the first and second folds compared to the others. The first fold also led to a bad segmentation mask for the esophagus, liver, and stomach. We want to highlight that the official BTCV webpage emphasizes that some patients may not have the right kidney or gallbladder and thus are not labelled; however, our network is capable of segmenting the right kidney independently of the folds, while the drop in performance in the second fold in the gallbladder may be related to the lack of annotated data.

\begin{table*}[t]
  \begin{center}
  \begin{adjustbox}{max width=1.0\textwidth}   
\begin{tabular}{l|rrrrrrrrrrrr|r}
\hline
Folds & SPL & RKI & LKI & GBL & ESO & LIV & STO & AOR & IVC & Veins & PAN & AG & \textbf{Avg.} \\
\hline
1 & 94.02 & 93.79 & 88.18 & 77.36 & 67.57 & 64.59 & 63.22 & 90.20 & 87.19 & 75.20 & 78.73 & 69.47 & 79.13 \\
2 & 96.42 & 92.90 & 94.60 & 54.10 & 76.13 & 97.02 & 81.59 & 93.39 & 86.04 & 76.75 & 71.81 & 69.08 & 82.49 \\
3 & 97.02 & 95.25 & 95.70 & 86.04 & 80.17 & 97.72 & 95.47 & 88.78 & 89.73 & 84.94 & 87.30 & 75.4 & 89.46 \\
4 & 48.38 & 88.81 & 92.57 & 91.69 & 79.01 & 95.06 & 91.35 & 89.43 & 86.65 & 70.76 & 78.05 & 75.89 & 82.3 \\
5 & 96.97 & 95.35 & 95.31 & 84.96 & 82.57 & 97.46 & 88.85 & 87.49 & 87.44 & 83.02 & 83.82 & 77.19 & 88.36 \\
\hline
\textbf{Avg} & 86.56 & 93.22 & 93.27 & 78.83 & 77.09 & 90.37 & 84.10 & 89.86 & 87.41 & 78.13 &79.94 &73.41 & 84.35 \\
\hline
\end{tabular}
\end{adjustbox}
\end{center}
\caption{The class-wise Dice scores, expressed in percentages, for each fold of SSH-UNet trained on the BTCV dataset using 5-fold cross-validation. Note: spleen (SPL), right kidney (RKI), left kidney (LKI), gallbladder (GBL), esophagus (ESO), liver (LIV), stomach (STO), aorta (AOR), inferior vena cava (IVC), portal and splenic veins (Venis), pancreas (PAN), left and right adrenal glands (AG).}
\label{tab:per_organ_kfold}
\end{table*}
\section{Ablation study}
\label{sec:ablation}
\subsection{Model components}
\begin{figure*}[t]
\begin{center}
\includegraphics[width=1.0\linewidth]{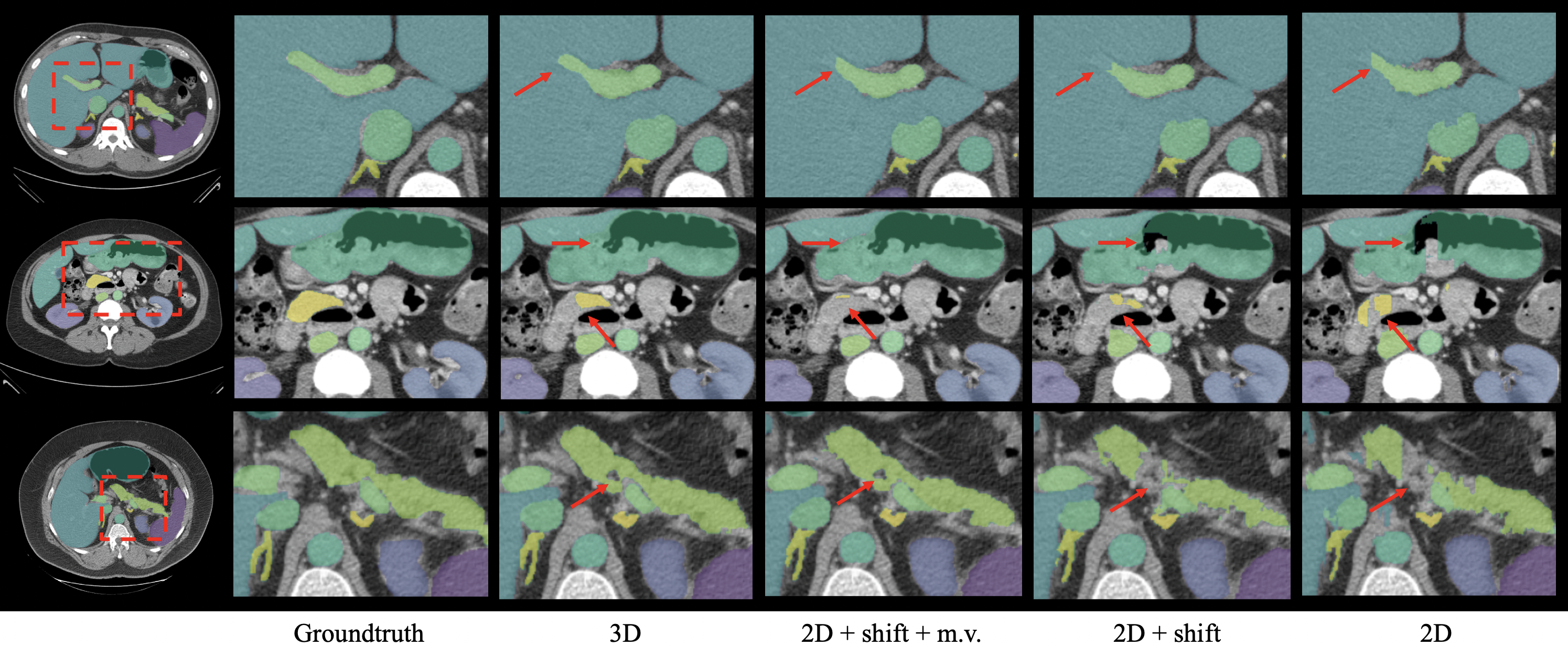}
\end{center}
   \caption {Qualitative results with representative samples from the BTCV dataset. The first row highlights the segmentation results for the portal and splenic veins. Fully 3D UNet achieves qualitatively the best result, while we can observe that the last three columns miss the segmentation of a small left portion. The second row focuses on the segmentation of the pancreas (yellow) and stomach (green). We can see that the 2D implementation of UNet and the 2D UNet with the shift operation (last two columns) are not able to segment a portion of the stomach, while our network (third column) and 3D UNet can perfectly segment it. In the last row, the pancreas is pointed again. In this case, it is segmented properly by both 3D UNet and our implementation while a portion is completely missed by the fully 2D model even if integrated with the shift operation.}
\label{fig:ablation_results}
\end{figure*}

We perform an ablation study to validate the effectiveness of the individual components of our model. As shown in Tables \ref{tab:ablation}, we can see the results of the different configurations trained with the BTCV and AMOS datasets. A UNet with only 2D convolution resulted in the lowest mDSC score. By introducing only the shift operation, referred to as "shift" in the table,  performance improved compared to the simple 2D case. With less than half of the parameters by combining multi-view with the shift operation (m.v. + shift) we are able to achieve comparable results of fully 3D UNet with the same architecture. In Figure \ref{fig:ablation_results} we can see qualitative results on the BTCV validation set.
\begin{table}[t]
  \begin{center}
    {\small{
\begin{tabular}{rrrr}
\hline
Components & Params & mDSC$_{BTCV}$ & mDSC$_{AMOS}$ \\
\hline
3D & 16.54 M & 0.842 & 0.882\\
2D & 6.18 M & 0.801 & 0.811\\
2D$+$shift &  6.48 M &  0.822 & 0.871\\
\hline
2D$+$shift$+$m.v.&  6.48 M &  0.838 & 0.873\\
\hline
\end{tabular}
}}
\end{center}
\caption{Ablation analysis of the introduced components. The term \textit{shift} stands for the slice shift operation, while the term \textit{m.v.} stands for multi-view fusion. In the last two columns, we can see the average Dice score on the validation set for BTCV and AMOS datasets.}
\label{tab:ablation}
\end{table}

\subsection{Shift operation}
We investigate the impact on the performance of the proportion of shifted channels. In Table \ref{tab:shifting_ablation} we can see that by shifting $1/4$ of the feature maps forwards and $1/4$ backwards (meaning we are shifting in total half of the channels) we have the best result. In the last column, we have the 2D case without shifting.
\begin{table}[t]
  \begin{center}
    {\small{
\begin{tabular}{r|rrrrr}
\hline
Shifted channels &$1/2$ & $1/4$ & $1/8$ & $1/16$ & $0$\\
\hline
mDSC & 0.866 & \textbf{0.871} & 0.868 & 0.865 & 0.811 \\
\hline
\end{tabular}
}}
\end{center}
\caption{Performance comparison on the proportion of channels shifted forward and backwards on the AMOS-CT validation set. Proportion $0$ is the fully 2D case without shift. The proportion $1/2$ is the case where all the channels are shifted, half forward and half backwards.}
\label{tab:shifting_ablation}
\end{table}

\subsection{Model complexity}
\begin{figure}[t]
\begin{center}
\includegraphics[width=1.0\linewidth]{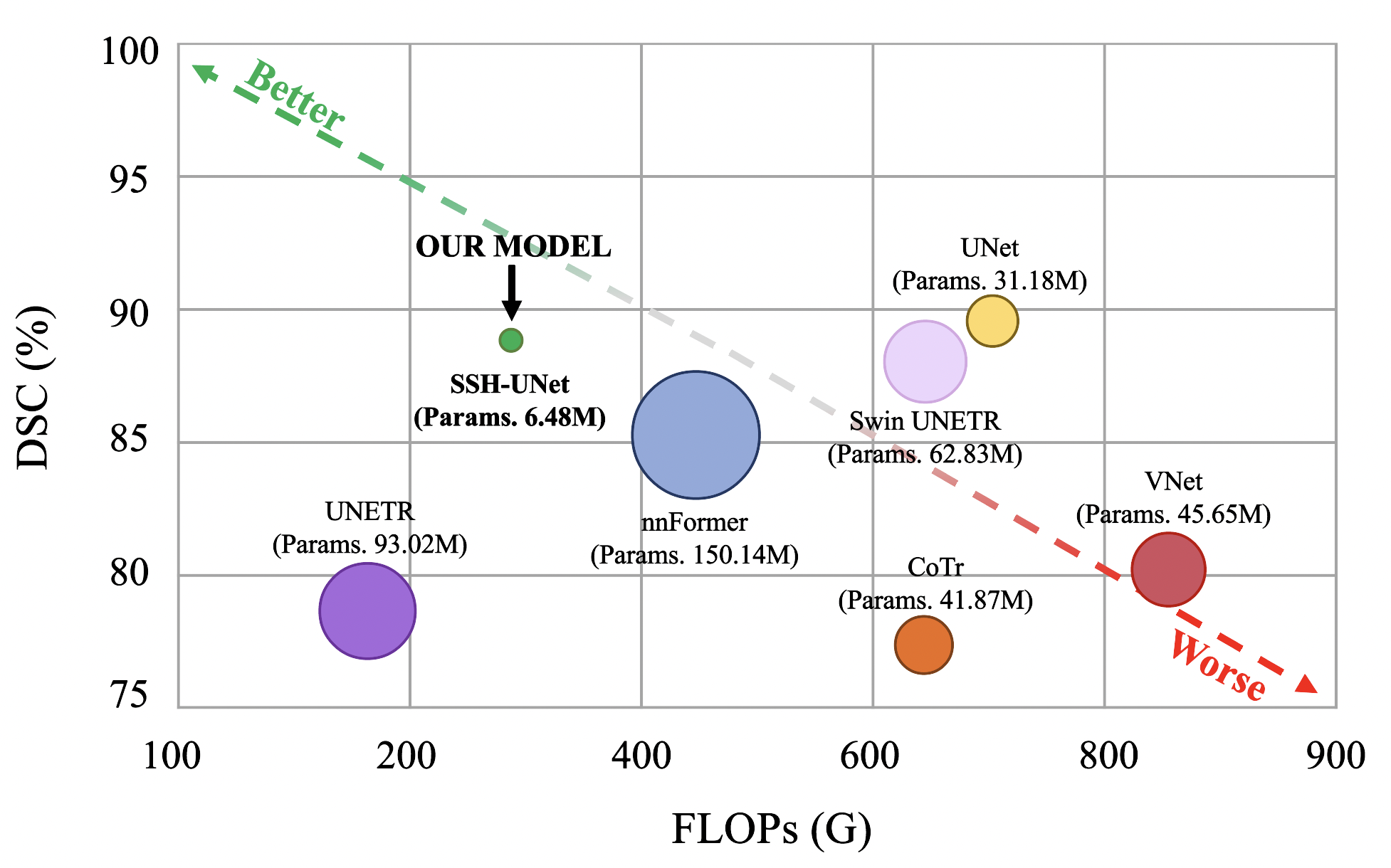}
\end{center}
   \caption{\textbf{Efficiency: FLOPs vs. DSC.} We plot the average DSC on the validation set of AMOS-CT. The FLOPs and parameters are estimated using $[1\times 128\times 128\times 128]$ as model input. The size of each circle indicates the number of parameters (Params.). We can observe that SSH-UNet has the lowest number of parameters and small FLOPs compared to other implementations while maintaining the second-highest DSC of 87.28\%.}
\label{fig:complexity}
\end{figure}

In this section, we examine the model complexity. In Table \ref{tab:complexity} the floating-point operations per second (FLOPs) and the number of parameters are presented for SSH-UNet and other baselines. A graphical representation of the Table can be seen in Figure \ref{fig:complexity}, where the efficiency plot shows that SSH-UNet is computationally more efficient compared with other state-of-the-art models (on average less than $1/5$ of parameters) while maintaining the second-highest DSC score of \textbf{87.28\%}.
\begin{table}[t]
  \begin{center}
    {\small{
\begin{tabular}{l|rr|rr}
\hline
Models & mDSC(\%)& Params(M) &Flops(G) \\
\hline
UNet & 88.87  & 31.18 & 680.31\\
VNet & 81.96  & 45.65 & 849.96\\
CoTr &  77.13 & 41.87 & 668.15\\
nnFormer & 85.63 & 150.14 & 425.78\\
UNETR &  78.33 & 93.02 & 177.51\\
Swin.UNETR & 86.37 & 62.83 &668.15\\
\hline
SSH-UNet (Ours) & 87.28 & 6.48 & 288.99 \\
\hline
\end{tabular}
}}
\end{center}
\caption{Overall results of six state-of-the-art methods taken from the official AMOS-CT validation benchmark in~\cite{ji2022amos} and SSH-UNet.}
\label{tab:complexity}
\end{table}

%% file: 10_conclusion.tex
\section{Conclusions}
\label{sec:conclusions}
Organ segmentation is a fundamental task in the medical field. The volumetric data that characterize CT and MRI acquisitions make, however, the segmentation task computationally expensive. On the one hand, 2D CNNs provide a low latency solution unable to capture inter-slice information, on the other hand, 3D CNNs extract three-dimensional features at the price of high computation costs and risk of overfitting. Moreover, popular 2.5D multi-view fusion methods train three separate networks where the features of the orthogonal planes are learned independently, despite being part of the same volume. In SSH-UNet this is addressed by imposing weight sharing between convolutions so that only one network needs to be trained and multi-view features are collaboratively learned.
In this work, we introduced a novel approach for the segmentation of volumetric medical data. Inspired by works in the field of Video Action Recognition we interpret the slices of a volume as the frame of a video. Given a 2D backbone, to re-integrate the information between features belonging to adjacent slices we leverage the power of a shifting mechanism inspired by the TSM module. Spatio-temporal modeling, declined on pseudo-3D operators, despite being well-known in the Video Understanding field was never used before in the medical image analysis to extract and mingle multi-slice features. Our network, by using a 2D convolution with weight sharing mechanism and slice shift, can extract 3D features keeping low computational complexity.
In comparison to other popular state-of-the-art methods, SSH-UNet achieves an accuracy of \textbf{87.28\%} on the AMOS validation providing the smallest model in terms of parameters (6.48M) compared to the best network which has $+1.6\%$ improve in accuracy but $\times5$ increase in parameters.

%% file: 12_appendix.tex
\appendix
\label{sec:appendix}

%% file: _main.bbl
\begin{thebibliography}{10}\itemsep=-1pt

\bibitem{msd}
Michela Antonelli, Annika Reinke, Spyridon Bakas, Keyvan Farahani, Annette
  Kopp-Schneider, Bennett~A Landman, Geert Litjens, Bjoern Menze, Olaf
  Ronneberger, Ronald~M Summers, et~al.
\newblock The medical segmentation decathlon.
\newblock {\em Nature communications}, 13(1):4128, 2022.

\bibitem{cao2021swin}
Hu Cao, Yueyue Wang, Joy Chen, Dongsheng Jiang, Xiaopeng Zhang, Qi Tian, and
  Manning Wang.
\newblock Swin-unet: Unet-like pure transformer for medical image segmentation.
\newblock {\em arXiv preprint arXiv:2105.05537}, 2021.

\bibitem{chen2021transunet}
Jieneng Chen, Yongyi Lu, Qihang Yu, Xiangde Luo, Ehsan Adeli, Yan Wang, Le Lu,
  Alan~L Yuille, and Yuyin Zhou.
\newblock Transunet: Transformers make strong encoders for medical image
  segmentation.
\newblock {\em arXiv preprint arXiv:2102.04306}, 2021.

\bibitem{chen2019s3d}
Wei Chen, Boqiang Liu, Suting Peng, Jiawei Sun, and Xu Qiao.
\newblock S3d-unet: separable 3d u-net for brain tumor segmentation.
\newblock In {\em Brainlesion: Glioma, Multiple Sclerosis, Stroke and Traumatic
  Brain Injuries: 4th International Workshop, BrainLes 2018, Held in
  Conjunction with MICCAI 2018, Granada, Spain, September 16, 2018, Revised
  Selected Papers, Part II 4}, pages 358--368. Springer, 2019.

\bibitem{dosovitskiy2021an}
Alexey Dosovitskiy, Lucas Beyer, Alexander Kolesnikov, Dirk Weissenborn,
  Xiaohua Zhai, Thomas Unterthiner, Mostafa Dehghani, Matthias Minderer, Georg
  Heigold, Sylvain Gelly, Jakob Uszkoreit, and Neil Houlsby.
\newblock An image is worth 16x16 words: Transformers for image recognition at
  scale.
\newblock In {\em International Conference on Learning Representations}, 2021.

\bibitem{futrega2022optimized}
Micha{\l} Futrega, Alexandre Milesi, Micha{\l} Marcinkiewicz, and Pablo
  Ribalta.
\newblock Optimized u-net for brain tumor segmentation.
\newblock In {\em International MICCAI Brainlesion Workshop}, pages 15--29.
  Springer, 2022.

\bibitem{Hatamizadeh_2022}
A. Hatamizadeh, Y. Tang, V. Nath, D. Yang, A. Myronenko, B. Landman, H.~R.
  Roth, and D. Xu.
\newblock Unetr: Transformers for 3d medical image segmentation.
\newblock In {\em Proceedings of the IEEE/CVF Winter Conference on Applications
  of Computer Vision}, pages 574--584, January 2022.

\bibitem{Isensee_2021}
Fabian Isensee, Paul~F Jaeger, Simon~AA Kohl, Jens Petersen, and Klaus~H
  Maier-Hein.
\newblock nnu-net: a self-configuring method for deep learning-based biomedical
  image segmentation.
\newblock {\em Nature methods}, 18(2):203--211, 2021.

\bibitem{jang2022m3t}
Jinseong Jang and Dosik Hwang.
\newblock M3t: Three-dimensional medical image classifier using multi-plane and
  multi-slice transformer.
\newblock In {\em Proceedings of the IEEE/CVF Conference on Computer Vision and
  Pattern Recognition}, pages 20718--20729, 2022.

\bibitem{ji2022amos}
Yuanfeng Ji, Haotian Bai, Chongjian Ge, Jie Yang, Ye Zhu, Ruimao Zhang, Zhen
  Li, Lingyan Zhanng, Wanling Ma, Xiang Wan, et~al.
\newblock Amos: A large-scale abdominal multi-organ benchmark for versatile
  medical image segmentation.
\newblock {\em Advances in Neural Information Processing Systems},
  35:36722--36732, 2022.

\bibitem{BTCV_orig}
Bennett Landman, Zhoubing Xu, J Igelsias, Martin Styner, T Langerak, and Arno
  Klein.
\newblock Miccai multi-atlas labeling beyond the cranial vault--workshop and
  challenge.
\newblock In {\em Proc. MICCAI Multi-Atlas Labeling Beyond Cranial
  Vault—Workshop Challenge}, volume~5, page~12, 2015.

\bibitem{li2019collaborative}
Chao Li, Qiaoyong Zhong, Di Xie, and Shiliang Pu.
\newblock Collaborative spatiotemporal feature learning for video action
  recognition.
\newblock In {\em Proceedings of the IEEE/CVF Conference on Computer Vision and
  Pattern Recognition}, pages 7872--7881, 2019.

\bibitem{Lin_2019}
Ji Lin, Chuang Gan, and Song Han.
\newblock Tsm: Temporal shift module for efficient video understanding.
\newblock In {\em Proceedings of the IEEE/CVF International Conference on
  Computer Vision}, pages 7083--7093, 2019.

\bibitem{Liu_2021}
Ze Liu, Yutong Lin, Yue Cao, Han Hu, Yixuan Wei, Zheng Zhang, Stephen Lin, and
  Baining Guo.
\newblock Swin transformer: Hierarchical vision transformer using shifted
  windows.
\newblock In {\em Proceedings of the IEEE/CVF International Conference on
  Computer Vision}, pages 10012--10022, 2021.

\bibitem{Luo_2019}
Chenxu Luo and Alan~L Yuille.
\newblock Grouped spatial-temporal aggregation for efficient action
  recognition.
\newblock In {\em Proceedings of the IEEE/CVF International Conference on
  Computer Vision}, pages 5512--5521, 2019.

\bibitem{milletari2016v}
Fausto Milletari, Nassir Navab, and Seyed-Ahmad Ahmadi.
\newblock V-net: Fully convolutional neural networks for volumetric medical
  image segmentation.
\newblock In {\em 2016 fourth international conference on 3D vision (3DV)},
  pages 565--571. Ieee, 2016.

\bibitem{nikolov2018deep}
Stanislav Nikolov, Sam Blackwell, Alexei Zverovitch, Ruheena Mendes, Michelle
  Livne, Jeffrey De~Fauw, Yojan Patel, Clemens Meyer, Harry Askham, Bernardino
  Romera-Paredes, et~al.
\newblock Deep learning to achieve clinically applicable segmentation of head
  and neck anatomy for radiotherapy.
\newblock {\em arXiv preprint arXiv:1809.04430}, 2018.

\bibitem{prasoon2013deep}
Adhish Prasoon, Kersten Petersen, Christian Igel, Fran{\c{c}}ois Lauze, Erik
  Dam, and Mads Nielsen.
\newblock Deep feature learning for knee cartilage segmentation using a
  triplanar convolutional neural network.
\newblock In {\em Medical Image Computing and Computer-Assisted
  Intervention--MICCAI 2013: 16th International Conference, Nagoya, Japan,
  September 22-26, 2013, Proceedings, Part II 16}, pages 246--253. Springer,
  2013.

\bibitem{Ronneberger_2015}
Olaf Ronneberger, Philipp Fischer, and Thomas Brox.
\newblock U-net: Convolutional networks for biomedical image segmentation.
\newblock In {\em International Conference on Medical image computing and
  computer-assisted intervention}, pages 234--241. Springer, 2015.

\bibitem{roth2014new}
Holger~R Roth, Le Lu, Ari Seff, Kevin~M Cherry, Joanne Hoffman, Shijun Wang,
  Jiamin Liu, Evrim Turkbey, and Ronald~M Summers.
\newblock A new 2.5 d representation for lymph node detection using random sets
  of deep convolutional neural network observations.
\newblock In {\em Medical Image Computing and Computer-Assisted
  Intervention--MICCAI 2014: 17th International Conference, Boston, MA, USA,
  September 14-18, 2014, Proceedings, Part I 17}, pages 520--527. Springer,
  2014.

\bibitem{Sudhakaran_2020}
Swathikiran Sudhakaran, Sergio Escalera, and Oswald Lanz.
\newblock Gate-shift networks for video action recognition.
\newblock In {\em Proceedings of the IEEE/CVF Conference on Computer Vision and
  Pattern Recognition}, pages 1102--1111, 2020.

\bibitem{Sun_2020}
Shanlin Sun, Yang Liu, Narisu Bai, Hao Tang, Xuming Chen, Qian Huang, Yong Liu,
  and Xiaohui Xie.
\newblock Attentionanatomy: A unified framework for whole-body organs at risk
  segmentation using multiple partially annotated datasets.
\newblock In {\em 2020 IEEE 17th International Symposium on Biomedical
  Imaging}, pages 1--5. IEEE, 2020.

\bibitem{Tang_2019}
Hao Tang, Xuming Chen, Yang Liu, Zhipeng Lu, Junhua You, Mingzhou Yang, Shengyu
  Yao, Guoqi Zhao, Yi Xu, Tingfeng Chen, and \etal.
\newblock Clinically applicable deep learning framework for organs at risk
  delineation in ct images.
\newblock {\em Nature Machine Intelligence}, 1(10):480--491, 2019.

\bibitem{Tang_2022}
Yucheng Tang, Dong Yang, Wenqi Li, Holger~R Roth, Bennett Landman, Daguang Xu,
  Vishwesh Nath, and Ali Hatamizadeh.
\newblock Self-supervised pre-training of swin transformers for 3d medical
  image analysis.
\newblock In {\em Proceedings of the IEEE/CVF Conference on Computer Vision and
  Pattern Recognition}, pages 20730--20740, 2022.

\bibitem{Wang_2016}
Limin Wang, Yuanjun Xiong, Zhe Wang, Yu Qiao, Dahua Lin, Xiaoou Tang, and Luc
  Van~Gool.
\newblock Temporal segment networks: Towards good practices for deep action
  recognition.
\newblock In {\em European conference on computer vision}, pages 20--36.
  Springer, 2016.

\bibitem{Xie_2021}
Y. Xie, J. Zhang, C. Shen, and Y. Xia.
\newblock Cotr: Efficiently bridging cnn and transformer for 3d medical image
  segmentation.
\newblock {\em Medical Image Computing and Computer Assisted Intervention},
  pages 171--180, 2021.

\bibitem{yun2019improvement}
Jihye Yun, Jinkon Park, Donghoon Yu, Jaeyoun Yi, Minho Lee, Hee~Jun Park,
  June-Goo Lee, Joon~Beom Seo, and Namkug Kim.
\newblock Improvement of fully automated airway segmentation on volumetric
  computed tomographic images using a 2.5 dimensional convolutional neural net.
\newblock {\em Medical image analysis}, 51:13--20, 2019.

\bibitem{zhang2022bridging}
Yichi Zhang, Qingcheng Liao, Le Ding, and Jicong Zhang.
\newblock Bridging 2d and 3d segmentation networks for computation-efficient
  volumetric medical image segmentation: An empirical study of 2.5 d solutions.
\newblock {\em Computerized Medical Imaging and Graphics}, page 102088, 2022.

\bibitem{zhou2021nnformer}
Hong-Yu Zhou, Jiansen Guo, Yinghao Zhang, Lequan Yu, Liansheng Wang, and Yizhou
  Yu.
\newblock nnformer: Interleaved transformer for volumetric segmentation.
\newblock {\em arXiv preprint arXiv:2109.03201}, 2021.

\end{thebibliography}
